\documentclass[aps,showpacs,twocolumn,preprintnumbers]{revtex4}
\usepackage{bm}
\begin{document}


\count255=\time\divide\count255 by 60 \xdef\hourmin{\number\count255}
  \multiply\count255 by-60\advance\count255 by\time
 \xdef\hourmin{\hourmin:\ifnum\count255<10 0\fi\the\count255}

\newcommand{\xbf}[1]{\mbox{\boldmath $ #1 $}}

\newcommand{\sixj}[6]{\mbox{$\left\{ \begin{array}{ccc} {#1} & {#2} &
{#3} \\ {#4} & {#5} & {#6} \end{array} \right\}$}}

\newcommand{\threej}[6]{\mbox{$\left( \begin{array}{ccc} {#1} & {#2} &
{#3} \\ {#4} & {#5} & {#6} \end{array} \right)$}}

\newcommand{\clebsch}[6]{\mbox{$\left( \begin{array}{cc|c} {#1} & {#2} &
{#3} \\ {#4} & {#5} & {#6} \end{array} \right)$}}

\newcommand{\iso}[6]{\mbox{$\left( \begin{array}{cc||c} {#1} & {#2} &
{#3} \\ {#4} & {#5} & {#6} \end{array} \right)$}}

\title{Interplay of the Chiral and Large $N_c$ Limits in $\pi N$ Scattering}

\author{Thomas D. Cohen}
\email{cohen@physics.umd.edu}

\affiliation{Department of Physics, University of Maryland, College
Park, MD 20742-4111}

\author{Richard F. Lebed}
\email{Richard.Lebed@asu.edu}

\affiliation{Department of Physics and Astronomy, Arizona State
University, Tempe, AZ 85287-1504}

\date{August, 2006}

\begin{abstract}
Light-quark hadronic physics admits two useful systematic expansions,
the chiral and $1/N_c$ expansions.  Their respective limits do not
commute, making such cases where both expansions may be considered to
be especially interesting.  We first study $\pi N$ scattering lengths,
showing that (as expected for such soft-pion quantities) the chiral
expansion converges more rapidly than the $1/N_c$ expansion, although
the latter nevertheless continues to hold.  We also study the
Adler-Weisberger and Goldberger-Miyazawa-Oehme sum rules of $\pi N$
scattering, finding that both fail if the large $N_c$ limit is taken
prior to the chiral limit.
\end{abstract}


\maketitle

\section{Introduction}

The $1/N_c$ expansion of QCD has proven to be a useful qualitative and
semi-quantitative approach to hadronic physics.  Recently, a number of
model-independent properties of meson-baryon scattering and their
associated baryon resonances have been
derived~\cite{CL,CL1N,CLSU3penta,CL70,ItJt} using the contracted SU($2
N_f$) symmetry emergent from QCD in the large $N_c$ limit~\cite{DJM1}.
This general approach verifies that results~\cite{ANW,HEHW,MK,MP}
previously derived in the context of Skyrme-type models are in fact
general results of large $N_c$ QCD.  In assessing the usefulness of
this approach it is relevant to recall that in nature $N_c \! = \! 3$,
and depending upon the observable, the $1/N_c$ corrections can be
quite large: indeed, large enough in some cases to render the
expansion useless.  Accordingly, it is important to try to understand
{\em which\/} types of observables are particularly likely to suffer
from large $1/N_c$ corrections.

In this context let us note that QCD allows another useful expansion,
namely, about its chiral limit~\cite{chiPT}.  The chiral expansion is
essentially an expansion in $\{m_\pi, q\}/\Lambda$, where $q$ is the
relevant momentum transfer in the process and $\Lambda$ is a typical
hadronic scale (such as $4 \pi f_\pi$).  The power counting of the
chiral expansion limits its validity to processes with low momentum
transfer.

This paper focuses on the interplay of the chiral and large $N_c$
limits for processes involving $\pi N$ scattering.  Several {\it a
priori\/} reasons suggest that this interplay should be important in
understanding at least one regime in which the $1/N_c$ might be
expected to break down badly, the soft-pion limit.  First, each
expansion is taken around a different limit, and the convergence need
not be uniform: The ordering of limits can matter.  When this occurs,
at least one of the expansions {\em must\/} break down.  Second, the
large $N_c$ consistency conditions used to derive the
model-independent results are based upon unitarity in $\pi N$
scattering.  For a typical pion momentum and energy of $O(N_c^0)$,
unitarity in $\pi N$ scattering imposes a constraint: The scattering
amplitude cannot grow without bound and hence cannot be $O(N_c^1)$ at
large $N_c$~\cite{DJM1}.  However, consider such scattering near the
soft limit of zero momentum and zero pion mass.  In that regime
unitarity imposes no constraints; zero-momentum scattering can have an
arbitrarily large amplitude without violating unitarity.  However,
this is precisely the regime of relevance to chiral physics.  Thus,
nothing protects the convergence of the $1/N_c$ expansion in the
chiral regime.

This ability of the interplay between chiral and $1/N_c$ limits to
shed light on the convergence of the $1/N_c$ expansion is provided by
a number of examples in which the two expansions are known not to
commute.  In particular, it known that the coefficient of the leading
term nonanalytic in $m_\pi^2$ ($\propto m_q^1$) dependence differs if
one if one takes the chiral limit before the large $N_c$ limit or
after~\cite{Com}.  For example, the nucleon mass has a term
proportional to $m_\pi^3 \propto m_q^{3/2}$ as its leading nonanalytic
contribution.  Simply expanding around the the chiral limit, one finds
\begin{equation}
m_N^{\rm LNA} = -\frac{3 g_A^2 }{32 f_\pi^2 m_\pi^2 } \, ,
\label{LNA1}
\end{equation}
where LNA indicates leading nonanalytic contribution.  Taking $N_c$
large prior to the chiral limit, one finds
\begin{equation}
m_N^{\rm LNA} = -\frac{9 g_A^2 }{32 f_\pi^2 m_\pi^2 } \, ,
\label{LNA2}
\end{equation}
which differs from Eq.~(\ref{LNA1}) by a factor of 3.  This behavior
is ubiquitous; it is seen in all static nucleon quantities.

It is not hard to uncover the origin of this difference in the special
role played by the $\Delta$ at large $N_c$.  Recall that $\Delta
\equiv m_\Delta -m_N$ scales as $1/N_c$, implying that the $\Delta$ is
degenerate with the nucleon at large $N_c$.  For small $N_c$ values
(including 3) one has $m_\pi \! < \! \Delta$, and in this region one
can compute using a standard chiral expansion with pion and nucleon
intermediate states alone to obtain the leading nonanalytic
dependence.  However, as $\Delta \rightarrow 0$ the regime of validity
of the standard chiral expansion goes to zero.  Had one started with a
large $N_c$ expansion, then the $\Delta$ would be taken as degenerate
with the nucleon, and pion loops with $\Delta$ intermediate states
contribute to the nonanalytic behavior, yielding a different result.

While this behavior is indeed ubiquitous, it is generally hidden from
view.  Note that there is no direct way to isolate from data the
leading nonanalytic behavior of static quantities.  The only direct
way to access them is via lattice simulations of the relevant
quantities for varying quark mass values.  In contrast, the quantities
associated with $\pi N$ scattering studied in this paper are directly
obtainable via experiment.

We focus here on two issues.  The first is a description of the
constants that parametrize the threshold behavior (i.e., the
scattering lengths).  Their chiral behavior has been known for
decades~\cite{Peccei}; however, their relationship with large $N_c$
has not received the attention it deserves.  For convenience we use
the formalism of Ref.~\cite{Peccei}, which uses a chiral Lagrangian
formalism including an explicit $\Delta$, facilitating our comparison
with the large $N_c$ limit.  The second issue concerns sum rules
relating low-energy or static quantities to integrals of $\pi N$
scattering; we focus on the Adler-Weisberger and
Goldberger-Miyazawa-Oehme sum rules.

In Sec.~\ref{1Nsec} we exhibit compute the $1/N_c$ expansions of the
$S$ and $P$ waves for $\pi N$ scattering, while Sec.~\ref{scat} studies
the scattering lengths in the $1/N_c$ and chiral expansions.  The
famous $\pi N$ scattering sum rules are considered in
Sec.~\ref{sumrules}, and in Sec.~\ref{concl} we conclude.

\section{$\pi N$ Scattering in the $1/N_c$ Expansion} \label{1Nsec}

In Ref.~\cite{CL1N} the current authors derived a master expression
for linear relations among $\pi N$ scattering amplitudes, including
subleading corrections in the $1/N_c$ expansion.  The method is very
straightforward: As has been known since the 1980s~\cite{MMSU2}, the
leading-order [$O(N_c^0)]$ amplitudes satisfy the rule $I_t \! = \!
J_t$.  But it is equally true~\cite{KapMan} that amplitudes with $|I_t
\! - \! J_t| \! = \! n$ are at most $O(1/N_c^n)$.  Therefore, all
terms of $O(1/N_c)$ not simply proportional to the leading-order
${\cal J} \! \equiv \! I_t \! = \!  J_t$ terms have either $x \!
\equiv \! I_t \! = \! J_t \! - \! 1$ or $y \equiv I_t \! = \! J_t \! +
\! 1$.  Once this is taken into account, a distinct $O(N_c^0)$ reduced
amplitude $s$, determined by the underlying QCD dynamics, appears for
each structure:
\begin{widetext}
\begin{eqnarray}
\lefteqn{S_{LL^\prime R R^\prime I_s J_s} = \sum_{\cal J} \left[
\begin{array}{ccc} 1 & R^\prime & I_s\\ R & 1 & {\cal J}
\end{array} \right] \left[
\begin{array}{ccc} L^\prime & R^\prime & J_s \\ R & L & {\cal J} \end{array}
\right] s_{{\cal J} L L^\prime}^t} & & \nonumber \\ 
&~&
-\frac{1}{N_c}\sum_x \left[ \begin{array}{ccc} 1 & R^\prime &
I_s\\ R & 1 & x
\end{array} \right] \left[
\begin{array}{ccc} L^\prime & R^\prime & J_s \\ R & L & {x+1} \end{array}
\right] s_{x L L^\prime}^{t(+)}
- \frac{1}{N_c}\sum_y \left[
\begin{array}{ccc} 1 & R^\prime & I_s\\ R & 1 & y
\end{array} \right] \left[
\begin{array}{ccc} L^\prime & R^\prime & J_s \\ R & L & {y-1} \end{array}
\right] s_{y L L^\prime}^{t(-)}
+  O (1/N_c^2) \ . \label{MPplus}
\end{eqnarray}
\end{widetext}
This expression represents a spinless, nonstrange isovector meson
($\pi$) scattering from a nonstrange baryon of $I \! = \! J \! = R$
(for $N_c$ large, baryons in this ground-state band, including the $N$
and $\Delta$, are stable), through the $L^{th}$ partial wave into an
intermediate state of quantum numbers $I_s, \, J_s$.  Primes indicate
final-state quantum numbers.  In fact, the quantum numbers of the
mesons can be generalized~\cite{MMSU2,MMSU3,CLSU3penta,CL70,ItJt}, and
the $I_t \! = \! J_t$ rule also holds for 3-flavor
processes~\cite{ItJt}.  The square-bracketed quantities in this
expression are a trivial redefinition of the standard $6j$ symbols,
called $[6j]$ symbols in Ref.~\cite{CL1N}, used to compactify the
notation:
\begin{equation}
\left\{\begin{array}{ccc} a & b & e \\ c & d & f
\end{array} \right\} \equiv
\frac{(-1)^{-(b+d+e+f)}}{([a][b][c][d])^{1/4}}
\left[\begin{array}{ccc} a & b & e \\ c & d & f \end{array}
\right] \ ,
\end{equation}
where $[a] \! \equiv \! 2a \! + \! 1$ denotes a multiplicity factor.
In particular, the $[6j]$ symbols retain all the usual triangle rules
of their $6j$ counterparts.

Restricting to the case of $\pi N$ scattering ($R \! = R^\prime \! =
\! \frac 1 2$), ${\cal J}$ can only be 0 or 1, $x$ can only be 0, and
$y$ can only be 1; indeed, no new structures arise at $O(1/N_c^2)$,
although the reduced amplitudes $s$ themselves have $O(1/N_c)$ and
higher-order corrections.  Of course, arbitrarily high partial waves
are permitted since $L$, $L^\prime$, and $J_s$ have no upper bound,
but since we are interested in soft pions with the characteristic
threshold behavior $k^{2L}$, only the lowest partial waves are of
interest.  For this paper we limit to $S$- and $P$-wave amplitudes.
The latter are especially interesting because $\Delta$ intermediate
states appear there as resonant poles and contribute prominently;
indeed, we shall see that one cannot obtain the correct large $N_c$
scaling of $P$-wave scattering lengths unless cancellations between
intermediate $N$ and $\Delta$ states occur.  This result is strongly
reminiscent of the consistency condition approach~\cite{DJM1} of
imposing order-by-order unitarity in powers of $1/N_c$ in meson-baryon
scattering.

Using Eq.~(\ref{MPplus}), the expansions for the $S$ and $P$ waves
read
\begin{eqnarray}
S_{11} & = & s^t_{000} - \frac{1}{N_c} \frac{2}{\sqrt{6}}
s^{t(-)}_{100} \ , \nonumber \\
S_{31} & = & s^t_{000} + \frac{1}{N_c} \frac{1}{\sqrt{6}}
s^{t(-)}_{100} \ ,
\end{eqnarray}
\begin{eqnarray}
P_{11} & = & \left( s^t_{011} + \frac 2 3 s^t_{111} \right) -
\frac{1}{N_c} \frac{2}{\sqrt{6}} \left( s^{t(+)}_{011} +
s^{t(-)}_{111} \right) \ , \nonumber \\
P_{31} & = & \left( s^t_{011} - \frac 1 3 s^t_{111} \right) -
\frac{1}{N_c} \frac{2}{\sqrt{6}} \left( s^{t(+)}_{011} -
\frac 1 2  s^{t(-)}_{111} \right) \ , \nonumber \\
P_{13} & = & \left( s^t_{011} - \frac 1 3 s^t_{111} \right) +
\frac{1}{N_c} \frac{2}{\sqrt{6}} \left( \frac 1 2 s^{t(+)}_{011} -
s^{t(-)}_{111} \right) \ , \nonumber \\
P_{33} & = & \left( s^t_{011} + \frac 1 6 s^t_{111} \right) -
\frac{1}{N_c} \frac{1}{\sqrt{6}} \left( s^{t(+)}_{011} +
s^{t(-)}_{111} \right) \ . \nonumber
\end{eqnarray}
Inverting these equations yields
\begin{eqnarray}
s^t_{000} & = & \frac 1 3 \left( S_{11} + 2 S_{31} \right) \ ,
\nonumber \\
\frac{s^{t(-)}_{100}}{N_c} & = & - \sqrt{6} \left[ \frac 1 3 \left(
S_{11} - S_{31} \right) \right] \ , \label{reducedS}
\end{eqnarray}
\begin{eqnarray}
s^t_{011} & = & \frac 1 3 \left\{ \left[ \frac 1 3 \left( P_{11} +
2P_{31} \right) \right] + 2 \left[ \frac 1 3 \left( P_{13} + 2P_{33}
\right) \right] \right\} \ , \nonumber \\
s^t_{111} & = & 2 \left\{ \left[ \frac 1 3 \left( P_{11} - P_{31}
\right) \right] - \left[ \frac 1 3 \left( P_{13} - P_{33} \right)
\right] \right\} \ ,
\nonumber \\
\frac{s^{t(+)}_{011}}{N_c} & = & - \sqrt{\frac 2 3} \left\{ \left[
\frac 1 3 \left( P_{11} \! + 2P_{31} \right) \right] \! - \! \left[
\frac 1 3 \left( P_{13} \! + 2P_{33} \right) \right] \right\} ,
\nonumber \\
\frac{s^{t(-)}_{111}}{N_c} & = & - \sqrt{\frac 2 3} \left\{ \left[
\frac 1 3 \left( P_{11} \! - P_{31} \right) \right] + 2 \left[
\frac 1 3 \left( P_{13} \!  - P_{33} \right) \right] \right\} .
\nonumber \\ & & \label{reducedP}
\end{eqnarray}
As mentioned above, no new structures arise at $O(1/N_c^2)$;
therefore, Eqs.~(\ref{reducedS})-(\ref{reducedP}) are {\em exact}.
The $1/N_c$ expansion predicts each reduced amplitude $s$ to be
$O(N_c^0)$ [although, again, each $s$ has its own $O(1/N_c)$
corrections].  We see that the second of Eq.~(\ref{reducedS}) and the
last two of Eq.~(\ref{reducedP}) must be $O(1/N_c)$ at all energies.
While performing such a calculation at arbitrary energy would require
a mastery of nonperturbative QCD, it is nevertheless possible to
compute the amplitudes for soft pions using the technology of the
chiral Lagrangian; this is the task we consider in the next section.

\section{Scattering Lengths} \label{scat}

The calculation of $S$- and $P$-wave scattering lengths of $\pi N$
scattering, performed by Peccei nearly four decades ago~\cite{Peccei}
was one of the very first calculations carried out using the chiral
Lagrangian formalism.  Scattering lengths refer to behavior of the
amplitudes in the soft-pion limit; the scattering length corresponding
to the $L_{IJ}$ partial wave amplitude is defined the coefficient of
its threshold ($ \propto \! k^{2L}$) behavior, where $k$ is the c.m.\
momentum of the $\pi$.  In addition, it is conventional to multiply by
$m_\pi^{2L+1}$ to obtain a dimensionless quantity.

The lowest-order (tree-level) results for the $S$- and $P$-wave
scattering amplitudes, including both $N$ and $\Delta$ intermediate
states, are presented in Appendix~\ref{explicit},
Eqs.~(\ref{aSm})--(\ref{aP12pm}), as are the relations between the
couplings used in Ref.~\cite{Peccei} and the usual couplings $g_{\pi N
N}$ and $g_{\pi N \Delta}$, and the large $N_c$ scalings of all
relevant quantities.  It is our task in this section to expand the
scattering amplitude expressions given in Appendix~\ref{explicit} in
$1/N_c$, and demonstrate that the scalings found for the full
partial-wave amplitudes in Eqs.~(\ref{reducedS})--(\ref{reducedP}) are
supported by their threshold behaviors.  In particular, using the
notation introduced in Eq.~(\ref{isospin}), we wish to show that
\begin{eqnarray}
a^+_S , \ a^+_{P_{1/2}} \! \! + 2a^+_{P_{3/2}}, \ a^-_{P_{1/2}} \! \!
- a^-_{P_{3/2}} & = & O(N_c^0) \ , \nonumber \\
a^-_S, \  a^+_{P_{1/2}} \! \! - a^+_{P_{3/2}}, \ a^-_{P_{1/2}} \! \!
+2 a^-_{P_{3/2}} & = & O(1/N_c) \ . \label{Nscaling}
\end{eqnarray}
In this way, one may explore the nature of the combined chiral and
large $N_c$ limits.

Let us begin by considering the chiral limit ($m_\pi \! \to \! 0$)
independent of the $1/N_c$ limit.  Including the factors of $m_\pi$
that make the scattering amplitudes dimensionless, the generic scaling
is $O(m_\pi^2)$.  The exceptions, as one may check using
Eqs.~(\ref{aSm})--(\ref{aP12pm}), are the combinations
\begin{equation} \label{achiral}
a^+_S , \ a^+_{P_{1/2}} \! \! + 2a^+_{P_{3/2}} \
a^-_{P_{1/2}} \! \! - a^-_{P_{3/2}} , \
a^-_{P_{3/2}} \! \! + a^+_{P_{3/2}} = O(m_\pi^3) \ .
\end{equation}
The result $a^+_S \! = \! O(m_\pi^3)$ gives the celebrated
Weinberg-Tomozawa (WT) relation~\cite{WT},
\begin{equation} \label{WT}
a_S^{I = 1/2} = -2 a_S^{I = 3/2} = \frac{g_V^2}{4\pi f_\pi^2}
\frac{m_\pi^2}{1+m_\pi/M} + \, O \! \left( \frac{m_\pi^3}{M^3} \right) ,
\end{equation}
where $M \! \equiv m_N$, and the ratio $-2a_S^{I = 3/2} \! / \, a_S^{I
= 1/2}$ experimentally equals unity to within 5\%~\cite{GAK,SAID}.

Now we present the amplitudes in the $1/N_c$ expansion.  The
expressions are presented twice: first imposing just the well-known
$N$-$\Delta$ large $N_c$ mass degeneracy
\begin{equation}
\Delta \equiv m_\Delta - m_N = O(1/N_c) \, ,
\end{equation}
and then imposing Eq.~(\ref{greln}), $g_{\pi N \Delta}/g_{\pi N N} \!
= \! \frac{3}{2} [1 \!  + \!  O(1/N_c^2)]$.  The purpose is to show
the necessity of including the $\Delta$ degree of freedom in some of
the $P$-wave (but not $S$-wave) results to obtain the correct $1/N_c$
counting.

Starting with the $S$ waves,
\begin{eqnarray}
a^+_S & = & -\frac{m_\pi^3}{16 \pi M^3} \frac{g_{\pi NN}^2 + 3g_{\pi
N\Delta}^2} {1+m_\pi/M} + O \! \left( \frac{1}{N_c^2} \right)
\nonumber \\ & = & -\frac{31 m_\pi^3 g_{\pi NN}^2}{64\pi M^3 (1 +
m_\pi/M)} + O \! \left( \frac{1}{N_c^2} \right) \ ,
\label{aSpN}
\end{eqnarray}
The factor $1+m_\pi/M$ is retained because it is a source of known
$O(1/N_c)$ corrections.  The exhibited term scales as $N_c^0$.  Next,
\begin{widetext}
\begin{eqnarray}
a^-_S & = & \frac{m_\pi^2}{32 \pi (1+m_\pi/M)} \left[
\frac{4g_V^2}{f_\pi^2} + \frac{m_\pi^2}{M^4} (g_{\pi N N}^2
+g_{\pi N \Delta}^2) \right]
+ \, O \! \left( \frac{1}{N_c^3} \right) \nonumber \\
& = & \frac{m_\pi^2}{128 \pi (1+m_\pi/M)} \left[
\frac{16g_V^2}{f_\pi^2} + \frac{13 m_\pi^2}{M^4} g_{\pi N N}^2 \right]
+ \, O \! \left( \frac{1}{N_c^3} \right) \ ,
\label{aSmN}
\end{eqnarray}
\end{widetext}
where the exhibited terms scale as $1/N_c$.  We see that the leading
term in a strict chiral expansion [the $g_V^2$ term in
Eq.~(\ref{aSmN})], which gives rise to Eq.~(\ref{WT}), is actually
subleading in the $1/N_c$ expansion and comparable to another term in
that expansion; this is an excellent illustration of the
noncommutativity of the chiral and large $N_c$ limits.  On the other
hand, we confirm the formal results $a_S^+ \! = O(N_c^0)$ and
$a_S^- \! = O(1/N_c)$.

Before proceeding, a comment is in order.  While the general
phenomenon of the noncommutativity of the large $N_c$ and chiral
limits is quite reminiscent of the leading nonanalytic chiral behavior
of static nucleon properties, the underlying {\em mechanism\/} in this
case is rather different.  In the static nucleon case, the key issue
is the role of the $\Delta$.  In the case of $a^-_S$, it is the
quantum numbers in the $t$ channel; its leading chiral behavior comes
from a $t$-channel exchange with quantum numbers that are
scalar-isovector (i.e., like the time component of the $\rho$ meson),
which are suppressed at large $N_c$~\cite{DJM1}.

Which expansion works better experimentally?  Since we are merely
keeping track of $1/N_c$ factors but calculating in the soft-pion
limit, it stands to reason that results based on the chiral limit
should be much more accurate in this case (as we saw for the WT
relation).  In order to quantify this belief, one may numerically
compute the leading coefficients in Eqs.~(\ref{aSpN})--(\ref{aSmN}).
Despite the former being formally $O(N_c^1)$ larger than the latter,
up to a sign the two are numerically almost precisely equal ($\pm
0.081$, respectively), surely a coincidence.  The strict $1/N_c$
expansion correctly predicts $a_S^-$ but grossly overestimates
$a_S^+$; experimentally, $a_S^- \! = \!  +0.08676$, $a_S^+ \! = \!
-0.00135$~\cite{SAID}.

The expansions for the $P$-wave scattering lengths are
\begin{widetext}
\begin{equation}
a^+_{P_{1/2}} \! + 2a^+_{P_{3/2}}
= \frac{g_{\pi N \Delta}^2 m_\pi}{144\pi M^3(1+m_\pi/M)} [16M
\Delta - 17m_\pi^2]
+ \, O \! \left( \frac{1}{N_c} \right) \ , \label{aPfirst}
\end{equation}
\begin{equation}
a^-_{P_{1/2}} \! - a^-_{P_{3/2}}
= \frac{g_{\pi N \Delta}^2 m_\pi}{288\pi M^3 (1+m_\pi/M)} [-8M
\Delta + 7m_\pi^2] + \, O \! \left( \frac{1}{N_c}
\right) \ , \label{aP2nd}
\end{equation}
%
which are both $O(N_c^0)$ [and $O(m_\pi^3)$; see below].  The
numerical comparisons are $+0.221$ vs.\ $+0.543 + O(1/N_c)$ and
$+0.0695$ vs.\ $-0.138 + O(1/N_c)$, respectively.  The agreement is
apparently poor, and demonstrates that the strict leading-order result
does not dominate; clearly, the $O(1/N_c)$ corrections are
substantial, as including even a naive estimate of their magnitude
$[O(1/3)]$ indicates: The $1/N_c$ expansion here is numerically true,
but not very predictive.
\end{widetext}

Superficially, one finds the other two combinations to be $O(N_c^0)$
as well:
\begin{equation}
a^-_{P_{1/2}} \! + 2a^-_{P_{3/2}}
= -\frac{m_\pi^2 (9g_{\pi N N}^2 - 4g_{\pi N \Delta}^2)}
{72\pi M^2} + \, O \! \left( \frac{1}{N_c} \right) \ ,
\end{equation}
\begin{equation}
a^+_{P_{1/2}} \! - a^+_{P_{3/2}}
= -\frac{m_\pi^2 (9g_{\pi N N}^2 - 4g_{\pi N \Delta}^2)}
{72\pi M^2} + \, O \! \left( \frac{1}{N_c} \right) \ ,
\end{equation}
but using the relation Eq.~(\ref{greln}) between $g_{\pi N N}$ and
$g_{\pi N \Delta}$, in fact the combinations turn out to be
\begin{widetext}
\begin{equation}
a^-_{P_{1/2}} \! + 2a^-_{P_{3/2}}
= \frac{g_{\pi NN}^2}{64\pi M^4 (1+m_\pi/M)} \left\{ 8M \left[ M
m_\pi^2 \left( \frac{4 g_{\pi N \Delta}^2}{9 g_{\pi N N}^2} - 1
\right) - m_\pi^2 \Delta + M \Delta^2 \right] - m_\pi^4
\right\} + \! O \left( \frac{1}{N_c^2} \right) \, , \label{aPnlast}
\end{equation}
\begin{equation}
a^+_{P_{1/2}} \! - a^+_{P_{3/2}}
= \frac{g_{\pi NN}^2}{64\pi M^4 (1+m_\pi/M)} \left\{ 8M \left[ M
m_\pi^2 \left( \frac{4 g_{\pi N \Delta}^2}{9 g_{\pi N N}^2} - 1
\right) - m_\pi^2 \Delta + M \Delta^2 \right] + m_\pi^4
\right\} + \! O \left( \frac{1}{N_c^2}
\right) \, . \label{aPlast}
\end{equation}
\end{widetext}
Using the scaling rules previously noted, one observes that the term
explicitly given is actually $O(1/N_c)$ because each term inside the
braces is $O(N_c^0)$.  The scaling rules demanded by
Eq.~(\ref{Nscaling}) are satisfied.  In this case, the numerical
comparisons are $-0.153$ vs.\ $+0.570 + O(1/N_c)$ and $-0.178$ vs.\
$+0.571 + O(1/N_c)$, respectively.  Again, the agreement is
unimpressive---after all, not even the signs are correct---but one
does not require excessively large $O(1/N_c)$ corrections to bring
them into accord.

Note moreover that the expansions given in
Eqs.~(\ref{aPfirst})--(\ref{aPlast}) curiously appear to violate the
$m_\pi^2$ and $m_\pi^3$ scaling described in and above
Eq.~(\ref{achiral}).  This is yet another manifestation of the
noncommutativity of the chiral and large $N_c$ limits, and
specifically originates in the question of whether $\delta \!
\equiv \! m_\pi / \Delta \! \to \! 0$ (chiral limit) or $\infty$
(large $N_c$ limit).  As a purely phenomenological matter $\delta \!
\approx \! 1/2$, and one can consider a hybrid limit with $m_\pi$,
$1/N_c \! \to \! 0$ but $\delta$ finite.  An excellent illustration of
this fact is given by the unique combination that is both $O(m_\pi^3)$
[see Eq.~(\ref{achiral})] and $O(1/N_c)$ [see Eq.~(\ref{Nscaling})],
which as one notes is the difference of Eq.~(\ref{aPnlast}) and
(\ref{aPlast}).  Indeed, this combination turns out to be simply the
threshold difference between $P_{13}$ and $P_{31}$.  Working in the
hybrid expansion, one finds
\begin{widetext}
\begin{eqnarray}
\lefteqn{a^-_{P_{1/2}} \! + 2a^-_{P_{3/2}} \! - a^+_{P_{1/2}} \! +
a^+_{P_{3/2}}} \nonumber \\
& \equiv & a_{P13} - a_{P31} = \frac{m_\pi^3}{16\pi M (1+m_\pi/M)}
\left\{ \frac{g_V^2}{f_\pi^2} - \frac{m_\pi}{M^3}
\left[ g_{\pi NN}^2 \! - \frac{2}{9} \left(
\frac{3-\delta}{1+\delta} \right) g_{\pi N \Delta}^2 \right]
\right\} \! + O \! \left( \frac{m_\pi^3}{N_c^3} \right) \! + O \!
\left( \frac{m_\pi^4}{N_c^2} \right) . \label{gold}
\end{eqnarray}
\end{widetext}
The $g_V$ term is $O(m_\pi^3/N_c^2)$ and the axial terms are
$O(m_\pi^4/N_c)$, and are now treated as comparable.  The
noncommutativity is clearly seen in the coefficient multiplying
$g_{\pi N \Delta}^2$: The factor in parentheses is $+3$ in the chiral
limit and $-1$ in the large $N_c$ limit; the latter appears when
taking Eq.~(\ref{aPnlast}) minus Eq.~(\ref{aPlast}).  The numerical
comparison in this case is $+0.0256$ to $+0.0035$ (strict large $N_c$
limit), $+0.0066$ (strict chiral limit), $+0.0056$ (physical value of
$\delta$; note that this value is numerically much closer to the
chiral than large $N_c$ limit).  Again, the central value is not
spot-on, since the corrections as indicated in Eq.~(\ref{gold}) easily
account for the difference, but note that the experimental value of
$+0.0256$ is numerically much smaller than either of the other
$P$-wave $O(m_\pi^3)$ combinations,
Eqs.~(\ref{aPfirst})--(\ref{aP2nd}), an indication that the extra
$1/N_c$ suppression is still significant even in the soft-pion case.
One sees that the chiral limit dominates these quantities, but that
$1/N_c$ suppressions still persist.

\section{Pion Scattering Sum Rules} \label{sumrules}

In this section we discuss the interplay of the chiral and large $N_c$
limits in pion scattering sum rules.  We first discuss the
Adler-Weisberger (AW) sum rule~\cite{AWreln} and then the
Goldberger-Miyazawa-Oehme (GMO) sum rule~\cite{GMOreln}, both of which
raise interesting physical issues.

The interplay of chiral dynamics and large $N_c$ QCD for the AW sum
rule is qualitatively different from that of the WT relation.  The WT
relation is an approximate relation that becomes increasingly exact as
the chiral limit is approached.  In contrast, the AW sum rule is
exact; it depends only on current algebra, certain assumptions about
the analytic structure of the theory (in particular, on the
convergence of the appropriate dispersion relation), and the absence
of $\pi N$ bound states.  Since the AW relation is a sum rule, it
requires information about pion scattering at all energies and not
merely for soft pions.  In this sense it is not a ``chiral'' relation.
However, as will become clear in this section, an understanding of the
relation at large $N_c$ depends upon some subtle issues in chiral
physics.

The AW sum rule reads
\begin{equation}
g_A^2 -1 = \frac{ 2 f_\pi^2}{\pi} \int_{m_\pi}^\infty \frac{d
\omega}{\omega} \, \sqrt{\omega^2-m_\pi^2} \, \left
(\sigma_{\pi^+ p} - \sigma_{\pi^- p} \right ) \, ,
\label{AW}\end{equation}
where $\sigma_{\pi^{\pm} p}$ is the total cross-section for
$\pi^{\pm}$ scattering off a proton as a function of $\pi$ c.m.\
energy $\omega$.

As noted by a number of groups~\cite{InBron}, this relation
superficially contradicts Witten's large $N_c$ scaling
rules~\cite{Witten}.  The apparent contradiction arises since
\begin{equation}
g_A^2 = O(N_c^2) \, , \  f_\pi^2 = O(N_c^1) \, , \
\sigma_{\pi^{\pm} p} = O(N_c^0) \ ,
\end{equation}
which seems to imply that the left-hand side of the relation
scales as $N_c^2$ while the right-hand side scales as $N_c^1$.

The resolution of this issue was discussed in detail by
Broniowski~\cite{Bron}, namely, the special rule played by the
$\Delta$ resonance.  Since the $\Delta$ is a partner of the nucleon in
the contracted SU(2$N_f$) symmetry, it behaves differently than other
resonances; its mass is anomalously low compared to typical
resonances, since $\Delta \! \equiv m_\Delta \! - \! m_N \! = \!
O(1/N_c)$ rather than $O(N_c^0)$ as for a typical baryonic excited
state, and has an anomalously strong coupling to the $\pi N$ channel:
$g_{\pi N \Delta} \! = \!  O(N_c^{1/2})$ rather than the typical
$O(N_c^0)$.  The contribution of the $\Delta$ is anomalously large,
$O(N_c^2)$, and it dominates the AW sum rule at both leading order and
next-to-leading order in a $1/N_c$ expansion.

The $\Delta$ contribution to the sum rule can be computed~\cite{Bron}
in the chiral limit and in a $1/N_c$ expansion.  It is tractable
because the $\Delta$ is narrow: Its width scales as $1/N_c^2$ in the
chiral limit.  Using a narrow-width approximation, the $\Delta$
contributes an amount $(g_A^*)^2$ to the right-hand side of the sum
rule, where $g_A^*$ is given by
\begin{equation}
g_A^* \equiv \frac{2 g_{\pi N \Delta}}{3 g_{\pi N N} } g_A + O(1/N_c)
= g_A + O(1/N_c ) \, , \label{star}
\end{equation}
where the Goldberger-Treiman (GT) relation~(\ref{GT}) has been used to
relate pion couplings to $g_A$.  The second equality follows from
contracted SU(2$N_f$) symmetry, which fixes the ratio $g_{\pi N
\Delta}/g_{\pi N N}$ to be $\frac{3}{2}$ at leading and
next-to-leading order in $1/N_c$.  This allows the AW sum rule to be
rewritten in the form
\begin{equation}
( g_A^2-{g_A^*}^2 ) -1 = \frac{ 2 f_\pi^2}{\pi}
\int_{m_\pi}^\infty \frac{d \omega}{\omega} \,
\sqrt{\omega^2-m_\pi^2} \, \left (\overline{\sigma}_{\pi^+ p} -
\overline{\sigma}_{\pi^- p} \right ) , \label{AWII}
\end{equation}
where $\overline{\sigma}_{\pi^\pm p}$ indicates cross sections with
the $\Delta$ contribution removed.  The quantity $\left (
g_A^2-{g_A^*}^2 \right )$ is $O(N_c^0)$, so that the left-hand side is
no longer characteristically larger than the right-hand side.  Indeed,
the right-hand side receives contributions from
$\overline{\sigma}_{\pi^+ p}$ and $\overline{\sigma}_{\pi^- p}$, each
of which is $O(N_c^1)$ (due to the $f_\pi^2$).  Evidently, these two
contributions must cancel to leading order in $1/N_c$.  An analysis of
why this happens appears in Ref.~\cite{Bron}.

We prefer to recast this analysis slightly, using the language of
Refs.~\cite{CL,CL1N,CLSU3penta,CL70,ItJt}.  To begin with, one should
note that the total $\pi N$ cross-section can be related to the
imaginary part of a forward scattering amplitude via the optical
theorem.  Moreover, by isospin invariance $\overline{\sigma}_{\pi^- p}
\! = \! \overline{\sigma}_{\pi^+ n}$, while by rotational invariance
the total cross section (and hence the forward amplitude) must be
independent of the spin state of the nucleon.  From the perspective of
the nucleon, the scattering amplitude can be represented by some
operator that depends on the nucleon spin and isospin and
parametrically on the momentum transfer.  Thus
$(\overline{\sigma}_{\pi^+ p} \! - \!
\overline{\sigma}_{\pi^- p})$ acts on
the space of nucleon states as a scalar-isovector operator.  However,
contracted SU(2$N_f$) symmetry implies the $I \! = \! J$
rule~\cite{KapMan}, which states that matrix elements of operators
characterized by an isospin $I$ and angular momentum $J$ in nucleon
states scale according to $N_c^{-|I-J|}$ times the generic $N_c$
scaling obtained via Witten's $N_c$ counting (which, for scattering
amplitudes, gives $N_c^0$).  This in turn implies
\begin{equation}
\left (\overline{\sigma}_{\pi^+ p} - \overline{\sigma}_{\pi^- p}
\right ) = O(1/N_c) \ ,
\end{equation}
which is the required cancellation.  Note that this cancellation
occurs at the level of the integrand and not the integral.

Reference~\cite{Bron} also considered the right-hand side in terms of
contributions of baryon resonances, and noted that it is not obvious
one may legitimately describe the right-hand side in terms of baryon
resonances since their widths are $O(N_c^0)$.  However, if one were to
describe the right-hand in terms of baryon resonances, one would find
that for a cancellation to occur at all energies (as it does in the
integral), the masses of classes of different resonances must be
(nearly) degenerate and their relative couplings must be proportional.
That is, large $N_c$ consistency rules must provide group-theoretic
constraints on the masses and couplings of the resonances.  It is
interesting to observe that that that formalism of
Refs.~\cite{CL,CL1N,CLSU3penta,CL70,ItJt} provides precisely such
constraints for baryon resonances.  At large $N_c$ they fall into
degenerate multiplets labeled by an emergent quantum number $K$, and
the relative couplings are fixed by SU(2$N_f$) Clebsch-Gordan factors.
Of course, this is hardly surprising: The underlying physical input of
the formalism of Refs.~\cite{CL,CL1N,CLSU3penta,CL70,ItJt} is the $I
\! = \! J$ rule for the $t$-channel amplitudes; the constraints emerge
when one recasts the physics into the $s$ channel.

The preceding analysis of this problem in Ref.~\cite{Bron} is quite
elegant and resolves the fundamental issues.  However this analysis
raises an interesting issue we wish to address here, concerning the
interplay of chiral dynamics and the large $N_c$ limit.  It seems
evident that this interplay is important, given the role of a chiral
relation (the GT relation) in deriving Eq.~(\ref{star}).  Indeed, as
noted above, Eq.~(\ref{star}) is strictly only valid in the chiral
limit; an incisive question is what happens away from this limit.  The
problem has an interesting feature, in that the AW sum rule is exact:
It has no chiral corrections.  Thus the matching of the two sides does
not depend upon the precise size of the explicit chiral symmetry
breaking; similarly the $N_c$ matching of the two sides should
continue to hold as one varies $m_\pi$.

In this context, the interplay of the chiral and large $N_c$ limits is
crucial.  If one takes the large $N_c$ limit prior to taking the
chiral limit, the $\Delta$ drops below $\pi N$ threshold and becomes
stable, contrary to the situation in the physical world.  If the
ordering of limits is taken the other way, the $\Delta$ remains above
threshold and accessible via $\pi N$ scattering.  The distinction
between the two cases is critical since it determines the region of
validity of the AW sum rule in Eq.~(\ref{AW}).  Recall that its
derivation requires the $\pi N$ scattering states to form a complete
set; thus, if there are $\pi N$ bound states the relation ceases to be
valid.  Thus if one takes the large $N_c$ limit prior to the chiral
limit, the AW sum rule is no longer valid, and for obvious reason: The
$\Delta$ contributes to the sum rule (indeed, dominates it) regardless
of whether or not it is above threshold.  However, the form of
Eq.~(\ref{AW}) is such that it contributes only if it is a scattering
state.  Thus, Eq.~(\ref{AW}) as written holds only if the $\Delta$ is
above threshold.

Suppose we restrict our attention to the regime of validity of
Eq.~(\ref{AW}), i.e., the region of an unstable $\Delta$.  Then, as
one approaches the large $N_c$ limit, of necessity one is approaching
the chiral limit:
\begin{equation}
m_\pi < \Delta = O(1/N_c) \, .
\end{equation}
This in turn implies that the size of chiral corrections are, of
necessity, bounded.  In particular, let us consider chiral corrections
to Eqs.~(\ref{star})--(\ref{AWII}).  These arise due to chiral
corrections to the GT relation, which essentially give a form factor
correction in the pion coupling from $q^2=0$ to $q^2=m_\pi^2$; as
such, it is of relative $O(m_\pi^2/\Lambda^2)$, where $\Lambda$ is a
typical hadronic scale characterizing the form factor.  In the regime
of validity of Eq.~(\ref{AW}), such a term is bounded parametrically
to be less than of relative $O(1/N_c^2)$.  Since $g_A^* \! = \!
O(N_c^1)$, this means that errors induced by chiral corrections are
bounded parametrically to be of $O(1/N_c)$ or less.  However, there
are already $1/N_c$ corrections of this order in Eq.~(\ref{star}).
Since corrections of this scale are consistent with (indeed, lead to)
the $N_c$ scaling of Eq.~(\ref{AWII}), it is apparent that chiral
corrections cannot spoil the $N_c$ counting consistency, provided the
system is in the regime of validity of the AW sum rule.

Next let us briefly consider the GMO sum rule.  This relation connects
a threshold property, $a_S^{-}$, to an integral over $\pi N$
scattering.  A useful representation is given in Ref.~\cite{Peccei}:
\begin{widetext}
\begin{equation}
  4 \pi  \left ( 1  +   \frac{m_\pi}{M} \right ) a_S^-  =
  \frac{2 g_{\pi N N}^2 m_\pi^2}{4 M^2 -m_\pi^2}  -
\frac{m_\pi^2}{\pi} \int_{m_\pi^2}^\infty \frac{d
\omega}{\sqrt{\omega^2-m_\pi^2}} \left (\sigma_{\pi^+ p} -
\sigma_{\pi^- p} \right ) \, , \label{GMO}
\end{equation}
\end{widetext}
where the first term on the right-hand side represents the
contribution of the nucleon Born terms.

As noted in Ref.~\cite{Bron}, many of the issues associated with the
large $N_c$ behavior of the GMO parallel the AW sum rule.  Like the AW
relation, there is an apparent discrepancy between the $N_c$ scaling
of the two sides.  The right-hand side naively scales as $N_c^1$ due
to the Born terms, while the left-hand side scales as $1/N_c$ from the
scaling of $a_S^-$.  Fortunately, the $\Delta$ contribution to the
scattering cancels off the nucleon Born term at both leading order and
the first subleading order in $1/N_c$, yielding consistency in $N_c$
counting.

As in the case of the AW sum rule, the GMO sum rule is exact provided
there are no bound states in the $\pi N$ channel, which constrains the
system to approach the chiral limit as fast as the large $N_c$ limit
in order to lie in the regime of validity of the sum rule.  This issue
is less important in the case of the GMO sum rule than for the AW sum
rule since the sum rule does not rely on current algebra, and hence
the scaling of the GT discrepancy with $m_\pi$ plays no role in the
derivation of the sum rule.

Similarly, the chiral behavior of the GMO sum rule is straightforward.
The left-hand side, as seen in the last section, is expected to be
$O(m_\pi^2)$, which matches the factor on the right-hand side.

However, the large $N_c$ and chiral behavior holds a surprise in the
chiral dependence of contributions proportional to $g_{\pi N N}^2$.
In Eq.~(\ref{GMO}) this term contributes at leading chiral order
[$O(m_\pi^2)$].  In contrast, in Eq.~(\ref{aSmN}) the term
proportional $g_{\pi N N}^2$ is of order $O(m_\pi^4)$.  Clearly, this
is possible only if the scattering term in the sum-rule cancels out
the dominant term proportional to $g_{\pi N N}^2$ and yielding a term
at a higher chiral order.  The scattering terms in the sum rule must
``know'' about the value of $g_{\pi N N}^2$ and conspire to cancel the
leading chiral behavior of the Born term.  This is a purely chiral
phenomenon and has nothing to do with $N_c$.  The way that this comes
about seems rather mysterious.  However, the $1/N_c$ expansion
provides some insights about how it can occur.  The dominant term
proportional to $g_{\pi N N}^2$ in the GMO sum rule is of order
$N_c^1$, two orders larger than the total.  Accordingly it is
canceled by the $\Delta$ contribution as it must be to ensure large
$N_c$ consistency .  Thus, the large $N_c$ limit provides a natural
mechanism by which the Born term may be canceled: exactly as is
needed to obtain consistency in chiral counting.

\section{Conclusions} \label{concl}

We have shown that the chiral and $1/N_c$ expansions may be considered
simultaneously, although one must take special care to accommodate the
noncommutativity of their limits associated with the parameter $\delta
\! \equiv \! m_\pi/(m_\Delta - m_N)$.

The scattering lengths, which are intrinsically soft-pion quantities,
naturally obey a more rapidly convergent chiral expansion than the
$1/N_c$ expansion; nevertheless, a strict $1/N_c$ expansion is still
meaningful, with numerically large but parametrically natural
$O(1/N_c)$ corrections.  We have also seen that a combination of
scattering amplitudes ($a_{P13} \! - \! a_{P31}$) at the same order of
$m_\pi$ but suppressed by $1/N_c$ is indeed experimentally smaller
than those without the $1/N_c$ suppression.  The $1/N_c$ expansion
continues to work even in the hostile environment of soft-pion
physics.

The Adler-Weisberger and Goldberger-Miyazawa-Oehme sum rules have an
especially interesting structure in terms of the chiral and $1/N_c$
expansions.  In particular, the $\Delta$ contribution must be
separated out of the cross section integrals by hand before the proper
counting in each expansion is manifest.  If one takes the $1/N_c$
limit prior to the chiral limit, then the $\Delta$ becomes a stable
state degenerate with the nucleon, and in particular is no longer a
$\pi N$ scattering state that appears in the integrals, invalidating
the usual forms of the sum rules.

{\it Acknowledgments.}  T.D.C.\ was supported by the D.O.E.\ through
grant DE-FGO2-93ER-40762; R.F.L.\ was supported by the N.S.F.\ through
grants PHY-0140362 and PHY-0456520.  We also thank Ron Workman for
information from the SAID Program, and Dan Dakin, who was involved in
the first stages of this project.

\appendix
\section{$\pi N$ Scattering Amplitude Expressions in Chiral
Perturbation Theory} \label{explicit}

For convenience of the reader, we present here the expressions for
$\pi N$ scattering from Ref.~\cite{Peccei}.  The calculation there was
presented at lowest (tree) order, and included $\pi$, $N$, $\rho$, and
$\Delta$ degrees of freedom; however, Ref.~\cite{Peccei} conveniently
describes how to eliminate from those expressions (in the modern
terminology, integrate out) $\rho$ degrees of freedom in favor of
$\pi$ only.

Three $\pi$-baryon couplings appear in Ref.~\cite{Peccei}: $f$ and
$f_0$ for $\pi N N$ in Eq.~(6), and $h$ for $\pi N \Delta$ in
Eq.~(17).  In perhaps more familiar notation,
\begin{eqnarray}
f   & = & -\frac{g_{\pi N N} \, m_\pi}{2m_N} \ , \nonumber \\
f_0 & = &  \frac{g_V \, m_\pi}{2f_\pi}   \ , \nonumber \\
h   & = &  \frac{g_{\pi N \Delta} \, m_\pi}{4\sqrt{2} m_N} \ ,
\end{eqnarray}
where numerically $g_{\pi N N} \! \simeq \! 13.5$ and $f_\pi \! \simeq
\!  93$~MeV, and $g_V \! = \! 1$ because it represents the polar
vector current coupling via the chirally-covariant derivative in the
$\pi N N$ kinetic energy term.  $g_{\pi N \Delta}$ can be extracted
from the $\Delta \! \to \! \pi N$ width (and turns out experimentally
to be about 20); however, in the spirit of the $1/N_c$ expansion, the
$N$ and $\Delta$ lie in the same multiplet and the ratio $g_{\pi N
\Delta}/g_{\pi N N}$ has a fixed value.  Indeed, if one uses the group
theory treating $N$ and $\Delta$ as bound states of $N_c$ $u$ and $d$
quarks, one finds~\cite{KP}
\begin{equation} \label{greln}
\frac{g_{\pi N \Delta}}{g_{\pi N N}} = \frac 3 2 \frac{\sqrt{(N_c - 1)
(N_c + 5)}}{N_c+2} = \frac 3 2 + O \left( \frac{1}{N_c^2} \right) \ .
\end{equation}
The absence of an $O(1/N_c)$ correction is known also from the
contracted spin-flavor symmetry that relates the $N$ and $\Delta$ at
large $N_c$~\cite{DJM1}.  As for $g_{\pi N N}$, it may be expressed
using the Goldberger-Treiman relation~\cite{GTref}
\begin{equation} \label{GT}
g_{\pi N N} = g_A \frac{m_N}{f_\pi} \ ,
\end{equation}
up to chiral corrections.  Since $g_A \! = \! O(N_c^1)$
(experimentally 1.26, it equals $(N_c \! + \! 2)/3$ in the quark
picture~\cite{KP}) while $m_N \! = \!  O(N_c^1)$ and $f_\pi \! = \!
O(N_c^{1/2})$, we have $g_{\pi N N}, \, g_{\pi N \Delta} \! = \!
O(N_c^{3/2})$, and therefore $f, \, h \! = \!  O(N_c^{1/2})$ and $f_0
\! = \! O(N_c^0)$.  It follows that
\begin{equation}
\frac{h^2}{f^2} = \frac{9}{32} + O \left( \frac{1}{N_c^2} \right) \, ,
\ \frac{f_0^2}{f^2} = O \left( \frac{1}{N_c^2} \right) \, .
\end{equation}
Similarly, as is well known, the $O(N_c^1)$ $\Delta$ mass $m_\Delta$
($M^*$ in Ref.~\cite{Peccei}) is split from $m_N$ ($M$ in
Ref.~\cite{Peccei}) only at $O(1/N_c)$, while $m_\pi \! = \!
O(N_c^0)$.

The results of Ref.~\cite{Peccei} are presented in terms of
isospin-even and -odd combinations of the $I \! = \! \frac 1 2$ and
$\frac 3 2$ amplitudes ${\cal A}$ for each partial wave
[Eqs.~(20)--(21)]:
\begin{equation} \label{isospin}
{\cal A}^+ = \frac{1}{3} \left( {\cal A}_{1/2} + 2 {\cal
A}_{3/2} \right) \, , \ {\cal A}^- = \frac{1}{3} \left( {\cal
A}_{1/2} - {\cal A}_{3/2} \right) \, .
\end{equation}
Comparing to Eqs.~(\ref{reducedS})--(\ref{reducedP}) above, we see
that ${\cal A}^+$ and ${\cal A}^-$ correspond neatly to $I_t \! = \!
0$ and 1, respectively.

According to the definitions of Ref.~\cite{Peccei}, Eq.~(25), the
scattering lengths are strictly speaking not actually lengths (or
volumes, for $P$ waves), but are made dimensionless by virtue of
including appropriate powers of $m_\pi$.  The $S$-wave scattering
lengths $a_S$ (i.e., the amplitudes in the soft-pion limit) have only
$J \! = \! \frac 1 2$, while $P$-wave scattering lengths $a_P$ (i.e.,
the first derivative of the amplitudes with respect to the square of
the $\pi$ c.m.\ momentum in the soft-pion limit) allow both $J \! = \!
\frac 1 2$ and $\frac 3 2$.

We now present the expressions of Ref.~\cite{Peccei} for $a_S^{\pm}$,
$a^{\pm}_{P_{1/2}}$, and $a^{\pm}_{P_{3/2}}$, where the final
subscript indicates the value of $J$.  From Eqs.~(30), (38), (33), and
(39), respectively,
\begin{widetext}
\begin{equation}
a_S^- = \frac{1}{4\pi (1+m_\pi \! /M)}
\left( \frac{2m_\pi^2}{4M^2 - m_\pi^2} f^2 + 2f_0^2 +
\frac{4m_\pi^2}{M^{*2}} \, h^2 \right) \, , \label{aSm}
\end{equation}
\begin{equation}
a_S^+ = \frac{1}{4\pi (1+m_\pi \! /M)}
\left[ -\frac{4Mm_\pi}{4M^2 - m_\pi^2} f^2 - \frac{8(2M^*
\! + M) m_\pi}{M^{*2}} h^2 \right] \, , \label{aSp}
\end{equation}
\begin{eqnarray}
\lefteqn{a_{P_{3/2}}^- = \frac{1}{36\pi (1+m_\pi \! /M)}
\left\{ -\frac{6}{(1-m_\pi \! /2M)^2} f^2 + \left(
-\frac{16m_\pi}{M^* \! - (M+m_\pi)}
+ \frac{m_\pi}{M^{*2} \! - (M-m_\pi)^2} \right. \right. }
& & \nonumber \\
& & \left. \left. \times \left[ 8M^* \! + 4M \! - 12m_\pi \!
+ \frac{8}{3M^*} \left( -M^2 \! + 6Mm_\pi \! + 3m_\pi^2 \right)
+ \frac{4}{3M^{*2}} \left( M^3 \! + 11M^2
m_\pi \! - 9M m_\pi^2 \! - 3m_\pi^3 \right) \right] \right) h^2
\right\} , \label{aP32m}
\end{eqnarray}
\begin{eqnarray}
\lefteqn{a_{P_{3/2}}^+ = \frac{1}{36\pi (1+m_\pi \! /M)}
\left\{ \frac{6}{(1-m_\pi \! /2M)^2} f^2 + \left(
\frac{32m_\pi}{M^* \! - (M+m_\pi)}
+ \frac{2m_\pi}{M^{*2} \! - (M-m_\pi)^2} \right. \right. }
& & \nonumber \\
& & \left. \left. \times \left[ 8M^* \! + 4M \! - 12m_\pi \!
+ \frac{8}{3M^*} \left( -M^2 \!
+ 6Mm_\pi \! + 3m_\pi^2 \right)
+ \frac{4}{3M^{*2}} \left( M^3 \! + 11M^2
m_\pi \! - 9M m_\pi^2 \! - 3m_\pi^3 \right) \right] \right) h^2
\right\} , \label{aP32p}
\end{eqnarray}
while from Eq.~(35), applied to not only $-$ but also $+$ amplitudes,
(34) and (40), respectively, we have
\begin{equation}
a^{\pm}_{P_{1/2}} \! = a^{\pm}_{P_{3/2}} \! - \frac{m_\pi^2}{4M^2} \,
a^{\pm}_S - \frac{m_\pi}{16\pi M} R^{(\pm)} \ , \label{aP12pm}
\end{equation}
where [Eqs.~(34) and (40), respectively]
\begin{eqnarray}
R^{(-)} & = & -\frac{4m_\pi^2}{4M^2 \! - m_\pi^2} f^2 - 4f_0^2
+ \left\{ -\frac{\frac{4}{3} \left( M^{*2} \! - M^2 \! - m_\pi^2
\right)} {\left[ \left( M^{*} \! + M \right)^2 - \! m_\pi^2 \right]
\left[ \left( M^* \! - M \right)^2 - \! m_\pi^2 \right]}
\right. \nonumber \\
& & \hspace{1em} \times \left[ 6m_\pi^2 + M (8M^* \! + 12M) +
\frac{8M}{3M^*} \left( M^2 \! - 3m_\pi^2 \right)
- \frac{2}{3M^{*2}} \left( 2M^4 - M^2 m_\pi^2
+ 9m_\pi^4 \right) \right] \nonumber \\
& & \left. +
\frac{\frac{32}{3}M^2 m_\pi^2}{\left[ \left( M^{*}
\! + M \right)^2 - \! m_\pi^2 \right] \left[ \left( M^* \! - M
\right)^2 - \! m_\pi^2 \right]} \left[ 1 + \frac{4M}{3M^*} -
\frac{1}{3M^{*2}} \left( M^2 - 6m_\pi^2 \right) \right] \right\} h^2 \
,
\end{eqnarray}
\begin{eqnarray}
R^{(+)} & = & \frac{32M^3}{m_\pi (4M^2 - m_\pi^2)} f^2
+ \left\{ \left[   \frac{44}{3} M^2 - 2M^{*2} + \frac{16MM^*}{3} +
4m_\pi^2 + \frac{16M}{3M^*} \left( M^2 - m_\pi^2
\right) - \frac{2 \left( M^2 - m_\pi^2 \right)^2}{M^{*2}} \right]
\right.
\nonumber \\
& & \left. \times \frac{\frac{16}{3} M m_\pi}{\left[
\left( M^{*} \! + M \right)^2 - \! m_\pi^2 \right] \left[ \left( M^*
\! - M \right)^2 - \! m_\pi^2 \right]} \right\} h^2 \ .
\end{eqnarray}
\end{widetext}

\end{document}